\documentstyle[12pt]{article}
\begin{document}
\setcounter{page}{0}
\renewcommand\arraystretch{1.3}
\title{\mbox{} \\[-2cm]
{\footnotesize\hspace*{\fill} DFFCUL Preprint \\[-4mm]
\footnotesize\hspace*{\fill} January 1994} \\[1.0cm]
Exact Conformal Scalar Field Cosmologies\thanks{Work
supported in part by a grant from DFFCUL-JNICT/CERN
(Pro\-gra\-m STRIDE/FEDER JN.91.02)}}
\author{{\bf Jo\~ao P. Abreu}\thanks{E-MAIL:
fjpabreu@scosysv.cc.fc.ul.pt}
 \\ [0.4em] {\bf Paulo Crawford}\thanks{E-MAIL:crawford@risc1.cc.fc.ul.pt}
\\[0.25em] and \\ [0.25em] {\bf Jos\'e P. Mimoso}\thanks{E-MAIL:
fmimoso@risc1.cc.fc.ul.pt} \\ [0.4em]
Departamento de F\'{\i}sica, Universidade de Lisboa \\
Campo Grande, Ed. C1, piso 4, 1700 Lisboa \\
Portugal}

\date{}
\maketitle
\baselineskip 20pt
\begin{abstract}

New exact solutions of Einstein's gravity coupled to a self-interacting
conformal scalar field are derived in this work. Our approach extends a
solution-generating technique originally introduced by Bekenstein for
massless conformal scalar fields. Solutions are obtained for a
Friedmann-Robertson-Walker geometry both for the cases of zero and non-zero
curvatures, and a variety of interesting  features are found.
It is shown that one class of solutions tends asymptotically to a
power-law inflationary behaviour $S(t)\sim t^p$ with $p>1$,
while another class exhibits a late time approach to the
$S(t)\sim t$ behaviour of the coasting models. Bouncing
models which avoid an initial singularity are also obtained. A general
discussion of the asymptotic behaviour and of the possibility of occurrence
of inflation is provided.
\end{abstract}

\vspace{1.5cm}
\pagebreak

{\Large\bf 1 \hspace{0.4cm} Introduction}

\vspace{0.8cm}
The action describing the interaction between the gravitational field and
a scalar field is generally constructed using the minimal coupling principle.
Only such terms are included that give the curved generalization of the
flat Minkowski spacetime form of the equation of motion for the scalar
field, i.e., of the Klein-Gordon equation. Terms that contain the
spacetime curvature are usually disregarded.

Recently, a non-minimal coupling between the spacetime curvature and a
scalar field has attracted a great deal of attention as a possible
improvement of the minimal coupling case [1--10]. In principle, the
consideration of such an interaction allows one to take more properly into
account the influence that, in the very early Universe, the extremely high
value of the curvature potentially had in the dynamical behaviour of this
coupled system. This point of view seems reinforced when the low energy
limit of unified theories (e.g., Superstrings, Supergravity, Kaluza-Klein
theories,\ldots) is considered. Upon compactification to four spacetime
dimensions, the effective action obtained in those theories often exhibits
the aforementioned feature [11].

If one requires that such a non-minimal coupling should: (i) involve only
the scalar field and not its derivatives, and (ii) be characterised by a
dimensionless coupling parameter; one is led uniquely to the following
action [1,2,12]
%%%%%%%%%%%%%%%%%%%%%%%%%%%%%%%%%%%%%%%%%%%%%%%%%%%%%%%%%%%%%%%%%%%%%%%%%%%%%
\begin{equation}
S[g^{ab},\phi]=\int d^4x\sqrt{-g}\left[\frac{M^2_{Pl}}{16\pi}R-\frac{1}{2}
\xi\phi^2R-\frac{1}{2}\partial_a\phi\partial^a\phi-V(\phi)\right],
\end{equation}
%%%%%%%%%%%%%%%%%%%%%%%%%%%%%%%%%%%%%%%%%%%%%%%%%%%%%%%%%%%%%%%%%%%%%%%%%%%%%
where $g$ is the determinant of the metric $g_{ab}, M_{Pl}=G_{N}^{-1/2}$ is
the Planck mass, $R$ is the Ricci scalar curvature, $\phi$ is a real scalar
field singlet, $V(\phi)$ is a general (effective) potential expressing the
self-interaction of $\phi$ and  $\xi$ is a dimensionless parameter that
characterises the strength of this finite coupling between the scalar field
and the curvature scalar. Note that the $\frac{1}{2}\xi\phi^{2}R$  term in
the action (1) is the only conceivable local scalar coupling satisfying the
requirements (i), (ii) stated above [12]. (Our sign conventions are that
$g_{ab}$ has signature $(-,+,+,+)$ and the Riemann and Ricci tensors are
defined as $R^{a}_{bcd}=\partial_{c}\Gamma^{a}_{bd}-\ldots,$
$R_{ab}=R^{c}_{acb}$. We use units such that $\hbar=c=1$, and usually we
shall set $8\pi/M^{2}_{Pl}=8\pi G_{N}=1$ for the sake of convenience.)

For $\xi=0$  we recover from eq.(1) the usual action for a minimally
coupled self-interacting scalar field. On the other hand, for
$\xi=\frac{1}{6}$ we obtain the conformal coupling case, and if $V(\phi)$
does not contain dimensional scales (i.e., the potential is zero or a
quartic power of $\phi$), it can be shown that the Klein-Gordon equation is
invariant under a conformal rescaling of the metric and the field itself,
i.e., is conformally invariant [13].

{}From expression (1), we find that its possible to define an effective
Newton gravitational constant $G^{eff}_{N}$, which is given by the inverse
of the coefficient of the curvature scalar, depending on the scalar field as
%%%%%%%%%%%%%%%%%%%%%%%%%%%%%%%%%%%%%%%%%%%%%%%%%%%%%%%%%%%%%%%%%%%%%%%%%%%%%
\begin{equation}
G^{eff}_{N}=\frac{1}{M^{2}_{Pl}-8\pi\phi^{2}}.
\end{equation}
%%%%%%%%%%%%%%%%%%%%%%%%%%%%%%%%%%%%%%%%%%%%%%%%%%%%%%%%%%%%%%%%%%%%%%%%%%%%%
Thus, if the scalar field is assumed homogeneous, the essential role of the
non-minimal coupling is to induce a time variation of the Newton
gravitational constant.

The theory based on (1) has been studied by a number of authors in several
different contexts. Hosotani [1], has discussed how the back-reaction of
gravity affects the stability of a scalar field, thus clarifying the
conditions for this theory to have an absolutely stable ground state in which
$\phi$ is constant and the spacetime background is maximally symmetric (in
connection with this issue see also Refs.\ [2] and [3]). Dolgov and Ford [4],
in turn, have investigated the possibility of a damping of the cosmological
constant, in an attempt to solve the cosmological constant
problem. In Ref. [5] the canonical quantization procedure is applied to (1)
and a homogeneous and isotropic minisuperspace model is discussed. Finally,
the prospects and consequences of inflation within this theory   have been
considered in a variety of works [6--8]. However, there aren't many exact
solutions in this context [9,10,14].

Here, we aim at addressing this question in the framework of the conformal
coupling case. We obtain new exact solutions of the coupled
Einstein--self-interacting conformal scalar field equations in the
particular case of a spatially homogeneous field propagating both in flat
and curved Friedmann-Robertson-Walker (FRW) spacetime backgrounds. This
is of considerable interest to assess the cosmological implications of this
theory and is done by extending a solution-generating technique originally
introduced by Bekenstein for massless conformal scalar fields, which is based
on the use of a conformal transformation [14]. This device, discussed at
length in Refs. [8,14,15], enables one to obtain a representation of the
theory in which the original coupled system is described by Einstein's
gravity plus a redefined \mbox{self-interacting} scalar field, which is now
a minimally coupled one. This new frame, sometimes called the Einstein frame,
is particularly suitable to the search of new exact solutions (e.g.,
following the strategy recently adopted by Ellis and Madsen [16]). Given a
solution in this new frame, the extended solution-generating technique
allows one to get a solution in the physical frame.

This paper is organized as follows. In the next Section, we derive and
review the general field equations for the non-minimal coupling model
characterised by the action (1). We also present here the equations that
govern the evolution of a FRW spacetime in the presence of a homogeneous
self-interacting conformal scalar field. In Section 3, we present the
extended solution-generating technique, reviewing in some detail the
conformal transformation upon which it is build and emphasizing its
usefulness regarding the derivation of exact conformal scalar field
cosmologies. In Section 4, we apply the extended solution-generating
technique to a number of classes of solutions in the Einstein frame that
were recently obtained by Ellis and Madsen [16]. The classes of solutions
obtained in the physical frame exhibit a diverse range of properties. For
instance, one class of solutions tends asymptotically to a power-law
inflationary behaviour, while another class exhibits a late time approach
to the $S(t)\sim t$ behaviour of the coasting models. Bouncing models
which avoid an initial singularity are also presented. Finally, we
conclude with a Section devoted to an overall summary and discussion of
the results.

\vspace{1.0cm}

{\Large\bf 2 \hspace{0.4cm} The Field Equations}

\vspace{0.8cm}

We begin this section  by deriving and reviewing the general field equations
for the non-minimal coupling model characterised by the action (1).

Variation of the action (1) with respect to the gravitational degrees of
freedom yields Einstein's equations
\begin{equation}
%%%%%%%%%%%%%%%%%%%%%%%%%%%%%%%%%%%%%%%%%%%%%%%%%%%%%%%%%%%%%%%%%%%%%%%%%%%%%
R_{ab}-\frac{1}{2}g_{ab}R=T_{ab},
\end{equation}
%%%%%%%%%%%%%%%%%%%%%%%%%%%%%%%%%%%%%%%%%%%%%%%%%%%%%%%%%%%%%%%%%%%%%%%%%%%%%
where $R_{ab}$ and $T_{ab}$  are the usual Ricci and energy-momentum tensors.
The later has the form
%%%%%%%%%%%%%%%%%%%%%%%%%%%%%%%%%%%%%%%%%%%%%%%%%%%%%%%%%%%%%%%%%%%%%%%%%%%%%
\begin{eqnarray}
T_{ab} & = &(1-\xi\phi^{2})^{-1}[(1-2\xi)\partial_{a}\phi\partial_{b}
\phi+(2\xi-\frac{1}{2})g_{ab}\partial_{c}\phi\partial^{c}\phi-
2\xi\phi\nabla_{a}(\partial_{b}\phi)- \nonumber \\
& &  g_{ab}V(\phi)+2\xi g_{ab}\phi\Box\phi],
\end{eqnarray}
%%%%%%%%%%%%%%%%%%%%%%%%%%%%%%%%%%%%%%%%%%%%%%%%%%%%%%%%%%%%%%%%%%%%%%%%%%%%%
with $\nabla_{a}$ standing for the covariant spacetime derivative associated
with the metric $g_{ab}$ and $\Box=\nabla_{a}\nabla^{a}$ is the corresponding
curved spacetime D'Alembertian. Variation  with respect to the scalar field
results in the Klein-Gordon equation
%%%%%%%%%%%%%%%%%%%%%%%%%%%%%%%%%%%%%%%%%%%%%%%%%%%%%%%%%%%%%%%%%%%%%%%%%%%%%
\begin{equation}
\Box\phi-\xi\phi R-V^{\prime}(\phi)=0,
\end{equation}
%%%%%%%%%%%%%%%%%%%%%%%%%%%%%%%%%%%%%%%%%%%%%%%%%%%%%%%%%%%%%%%%%%%%%%%%%%%%%
where $V^{\prime}(\phi)=dV(\phi)/d\phi$.

Next, we specialize the field equations for the particular case of a FRW
spacetime background filled with a homogeneous self-interacting conformal
scalar field.

The line element for a FRW spacetime can be written in standard normalized
comoving coordinates $(t,r,\theta,\varphi)$  as
%%%%%%%%%%%%%%%%%%%%%%%%%%%%%%%%%%%%%%%%%%%%%%%%%%%%%%%%%%%%%%%%%%%%%%%%%%%%%
\begin{equation}
ds^{2}=-dt^{2}+S^{2}(t)\left(\frac{dr^{2}}{1-kr^{2}}+r^{2}(d\theta^{2}+\sin^
{2}\theta d\varphi^{2})\right),
\end{equation}
%%%%%%%%%%%%%%%%%%%%%%%%%%%%%%%%%%%%%%%%%%%%%%%%%%%%%%%%%%%%%%%%%%%%%%%%%%%%%
where $S(t)$  represents the cosmic scale factor and $k$ is the curvature
index ($k=0,\pm1$ corresponds to the flat, closed and open FRW model,
respectively). The independent components of Einstein's equations (3)--(4)
for this metric can be represented by the Friedmann and Raychaudhuri
equations. In the conformal coupling case they become
%%%%%%%%%%%%%%%%%%%%%%%%%%%%%%%%%%%%%%%%%%%%%%%%%%%%%%%%%%%%%%%%%%%%%%%%%%%%%
\begin{equation}
3H^{2}+3K=(1-\zeta^{2}\phi^{2})^{-1}\left[3\zeta^{2}\dot{\phi}^{2}+V(\phi)+
\phi\dot{\phi}H\right],
\end{equation}
%%%%%%%%%%%%%%%%%%%%%%%%%%%%%%%%%%%%%%%%%%%%%%%%%%%%%%%%%%%%%%%%%%%%%%%%%%%%%
%%%%%%%%%%%%%%%%%%%%%%%%%%%%%%%%%%%%%%%%%%%%%%%%%%%%%%%%%%%%%%%%%%%%%%%%%%%%%
\begin{equation}
3\dot{H}+3H^{2}=(1-\zeta^{2}\phi^{2})^{-1}\left[V(\phi)-3\zeta^{2}
(\dot{\phi}^{2}-\phi\dot{\phi}H-\phi\ddot{\phi})\right],
\end{equation}
%%%%%%%%%%%%%%%%%%%%%%%%%%%%%%%%%%%%%%%%%%%%%%%%%%%%%%%%%%%%%%%%%%%%%%%%%%%%%
where, in accordance with the FRW symmetry, we have assumed that
$\phi=\phi(t)$. Here $K$ is the purely spatial part of the scalar curvature
$K=k/S^{2},\zeta=(\frac{1}{6})^{1/2}$ and, as usual,
$H=\dot{S}/S$ denotes the Hubble parameter and the overdot means
derivative with respect to the synchronous cosmic time, t. On the other hand,
using eqs.(7) and (8), the Klein-Gordon equation (5) becomes
%%%%%%%%%%%%%%%%%%%%%%%%%%%%%%%%%%%%%%%%%%%%%%%%%%%%%%%%%%%%%%%%%%%%%%%%%%%%%
\begin{equation}
\ddot{\phi}+3H\dot{\phi}+\zeta^{2}\phi(1-\zeta^{2}\phi^{2})^{-1}\left[
4V(\phi)+3\phi\dot{\phi}H+\phi\ddot{\phi}\,\right]+V^{\prime}(\phi)=0.
\end{equation}
%%%%%%%%%%%%%%%%%%%%%%%%%%%%%%%%%%%%%%%%%%%%%%%%%%%%%%%%%%%%%%%%%%%%%%%%%%%%%

It is worthwhile to point here that, for the case of a homogeneous scalar
field propagating in a spatially homogeneous and isotropic spacetime
background, the energy-momentum tensor (4) can be rewritten in the form of a
perfect fluid for any value of $\xi$ [2,7]. The energy density and pressure
of the equivalent perfect fluid are, in the conformal coupling case,
given by
%%%%%%%%%%%%%%%%%%%%%%%%%%%%%%%%%%%%%%%%%%%%%%%%%%%%%%%%%%%%%%%%%%%%%%%%%%%%%
\begin{equation}
\rho_{\phi}=(1-\zeta^{2}\phi^{2})^{-1}[3\zeta^{2}\dot{\phi}^{2}+V(\phi)
+\phi\dot{\phi}H],
\end{equation}
%%%%%%%%%%%%%%%%%%%%%%%%%%%%%%%%%%%%%%%%%%%%%%%%%%%%%%%%%%%%%%%%%%%%%%%%%%%%%
and
%%%%%%%%%%%%%%%%%%%%%%%%%%%%%%%%%%%%%%%%%%%%%%%%%%%%%%%%%%%%%%%%%%%%%%%%%%%%%
\begin{equation}
p_{\phi}=(1-\zeta^{2}\phi^{2})^{-1}[\zeta^{2}(\dot{\phi}^{2}-2\phi\ddot{\phi}
-4\phi\dot{\phi}H)-V(\phi)],
\end{equation}
%%%%%%%%%%%%%%%%%%%%%%%%%%%%%%%%%%%%%%%%%%%%%%%%%%%%%%%%%%%%%%%%%%%%%%%%%%%%%
respectively. It is evident that neither the energy density nor the pressure
are sign-definite. As a consequence, $\frac{1}{2}(\rho_{\phi}+3p_{\phi})$
has indeterminate sign, so that the strong energy condition [17]
%%%%%%%%%%%%%%%%%%%%%%%%%%%%%%%%%%%%%%%%%%%%%%%%%%%%%%%%%%%%%%%%%%%%%%%%%%%%%
\begin{equation}
(T_{ab}-\frac{1}{2}Tg_{ab})U^{a}U^{b}=\frac{1}{2}(\rho_{\phi}+3p_{\phi})
\geq0,
\end{equation}
%%%%%%%%%%%%%%%%%%%%%%%%%%%%%%%%%%%%%%%%%%%%%%%%%%%%%%%%%%%%%%%%%%%%%%%%%%%%%
where $U^{a}=\partial^{a}\phi/\sqrt{-\partial_{b}\phi\partial^{b}\phi}$
denotes the 4-velocity vector of the equivalent perfect fluid,
may or may not be valid. In the later case, the Hawking-Penrose theorem on
the existence of singularities [18] is  not applicable and singularity-free
solutions to equations (7)--(9) are then possible. In Section 4 a number of
singularity-free solutions of those equations will be presented.

We will find it convenient to define a new scalar field variable $\psi$
such that $\psi=\zeta\phi$. Using this variable eqs.(7)--(9) become
respectively
%%%%%%%%%%%%%%%%%%%%%%%%%%%%%%%%%%%%%%%%%%%%%%%%%%%%%%%%%%%%%%%%%%%%%%%%%%%%%
\begin{equation}
3H^{2}+3K=(1-\psi^{2})^{-1}[3\dot{\psi}^{2}+V(\psi)+\zeta^{-2}\psi
\dot{\psi}H],
\end{equation}
%%%%%%%%%%%%%%%%%%%%%%%%%%%%%%%%%%%%%%%%%%%%%%%%%%%%%%%%%%%%%%%%%%%%%%%%%%%%%
%%%%%%%%%%%%%%%%%%%%%%%%%%%%%%%%%%%%%%%%%%%%%%%%%%%%%%%%%%%%%%%%%%%%%%%%%%%%%
\begin{equation}
3\dot{H}+3H^{2}=(1-\psi^{2})^{-1}[V(\psi)-3(\dot{\psi}^{2}-\psi\dot{\psi}H-
\psi\ddot{\psi})],
\end{equation}
%%%%%%%%%%%%%%%%%%%%%%%%%%%%%%%%%%%%%%%%%%%%%%%%%%%%%%%%%%%%%%%%%%%%%%%%%%%%%
%%%%%%%%%%%%%%%%%%%%%%%%%%%%%%%%%%%%%%%%%%%%%%%%%%%%%%%%%%%%%%%%%%%%%%%%%%%%%
\begin{equation}
\ddot{\psi}+3H\dot{\psi}+\psi(1-\psi^{2})^{-1}[4\zeta^{2}V(\psi)+3\psi
\dot{\psi}H+\psi\ddot{\psi}]+\zeta^{2}V^{\prime}(\psi)=0,
\end{equation}
%%%%%%%%%%%%%%%%%%%%%%%%%%%%%%%%%%%%%%%%%%%%%%%%%%%%%%%%%%%%%%%%%%%%%%%%%%%%%
where now $V^{\prime}(\psi)=dV(\psi)/d\psi$.

\vspace{1.0cm}

{\Large\bf 3 \hspace{0.4cm} The Extended Solution-Generating Technique}

\vspace{0.8cm}

As we remarked in the Introduction, the solution-generating technique
employed in this work is based on the use of a conformal transformation of
the spacetime metric, which reduces the theory under consideration to one
containing canonical Einstein's gravity plus a canonical scalar field. In
what follows we present this solution-generating technique, reviewing in
some detail the conformal transformation upon which it is build.

Let us perform the following Weyl conformal transformation
%%%%%%%%%%%%%%%%%%%%%%%%%%%%%%%%%%%%%%%%%%%%%%%%%%%%%%%%%%%%%%%%%%%%%%%%%%%%%
\begin{equation}
\tilde{g}_{ab}=\Omega^{2}g_{ab} \quad , \quad
\tilde{g}^{ab}=\Omega^{-2}g^{ab} \quad , \quad
\tilde g=\Omega^{8}g,
\end{equation}
%%%%%%%%%%%%%%%%%%%%%%%%%%%%%%%%%%%%%%%%%%%%%%%%%%%%%%%%%%%%%%%%%%%%%%%%%%%%%
where
%%%%%%%%%%%%%%%%%%%%%%%%%%%%%%%%%%%%%%%%%%%%%%%%%%%%%%%%%%%%%%%%%%%%%%%%%%%%%
\begin{equation}
\Omega^{2}=|1-\xi\phi^{2}|.
\end{equation}
%%%%%%%%%%%%%%%%%%%%%%%%%%%%%%%%%%%%%%%%%%%%%%%%%%%%%%%%%%%%%%%%%%%%%%%%%%%%%
Since the transformed scalar curvature is
%%%%%%%%%%%%%%%%%%%%%%%%%%%%%%%%%%%%%%%%%%%%%%%%%%%%%%%%%%%%%%%%%%%%%%%%%%%%%
\begin{equation}
\tilde R=\Omega^{-2}[R-6\Omega^{-1}\Box\Omega],
\end{equation}
%%%%%%%%%%%%%%%%%%%%%%%%%%%%%%%%%%%%%%%%%%%%%%%%%%%%%%%%%%%%%%%%%%%%%%%%%%%%%
it is then straightforward to show that action (1) can be
written in the Einstein frame as
%%%%%%%%%%%%%%%%%%%%%%%%%%%%%%%%%%%%%%%%%%%%%%%%%%%%%%%%%%%%%%%%%%%%%%%%%%%%%
\begin{equation}
S[\tilde g^{ab},\phi]=\int d^{4}x\sqrt{-\tilde g}\left[\frac{\tilde R}{2}-
\frac{1}{2}F^{2}(\phi)\partial_{a}\phi\partial^{a}\phi-\tilde V(\phi)\right],
\end{equation}
%%%%%%%%%%%%%%%%%%%%%%%%%%%%%%%%%%%%%%%%%%%%%%%%%%%%%%%%%%%%%%%%%%%%%%%%%%%%%
where
%%%%%%%%%%%%%%%%%%%%%%%%%%%%%%%%%%%%%%%%%%%%%%%%%%%%%%%%%%%%%%%%%%%%%%%%%%%%%
\begin{equation}
F^2(\phi)=\frac{1-\xi(1-6\xi)\phi^{2}}{(1-\xi\phi^{2})^{2}},
\end{equation}
%%%%%%%%%%%%%%%%%%%%%%%%%%%%%%%%%%%%%%%%%%%%%%%%%%%%%%%%%%%%%%%%%%%%%%%%%%%%%
and
%%%%%%%%%%%%%%%%%%%%%%%%%%%%%%%%%%%%%%%%%%%%%%%%%%%%%%%%%%%%%%%%%%%%%%%%%%%%%
\begin{equation}
\tilde V(\phi)=\frac{V(\phi)}{(1-\xi\phi^{2})^{2}}.
\end{equation}
%%%%%%%%%%%%%%%%%%%%%%%%%%%%%%%%%%%%%%%%%%%%%%%%%%%%%%%%%%%%%%%%%%%%%%%%%%%%%
In order to put the kinetic term of the scalar field in the
canonical form, we define a new field $\tilde \phi$ as
%%%%%%%%%%%%%%%%%%%%%%%%%%%%%%%%%%%%%%%%%%%%%%%%%%%%%%%%%%%%%%%%%%%%%%%%%%%%%
\begin{equation}
\tilde \phi=\int d\phi F(\phi).
\end{equation}
%%%%%%%%%%%%%%%%%%%%%%%%%%%%%%%%%%%%%%%%%%%%%%%%%%%%%%%%%%%%%%%%%%%%%%%%%%%%%
The theory can thus be written in terms of these new variables as a minimally
coupled theory with the action
%%%%%%%%%%%%%%%%%%%%%%%%%%%%%%%%%%%%%%%%%%%%%%%%%%%%%%%%%%%%%%%%%%%%%%%%%%%%%
\begin{equation}
S[\tilde g^{ab},\tilde \phi]=\int d^{4}x\sqrt{-\tilde g}\left[\frac{\tilde R}
{2}-\frac{1}{2}\partial_{a}\tilde \phi\partial^{a}\tilde \phi-\tilde
V(\tilde \phi)\right].
\end{equation}
%%%%%%%%%%%%%%%%%%%%%%%%%%%%%%%%%%%%%%%%%%%%%%%%%%%%%%%%%%%%%%%%%%%%%%%%%%%%%
Taking $\xi=\frac{1}{6}$ in eq.(22) and solving for $\phi$  yields
%%%%%%%%%%%%%%%%%%%%%%%%%%%%%%%%%%%%%%%%%%%%%%%%%%%%%%%%%%%%%%%%%%%%%%%%%%%%%
\begin{equation}
\phi(\tilde \phi)=\zeta^{-1}\tanh(\zeta\tilde \phi),
\end{equation}
%%%%%%%%%%%%%%%%%%%%%%%%%%%%%%%%%%%%%%%%%%%%%%%%%%%%%%%%%%%%%%%%%%%%%%%%%%%%%
and upon substitution of this result in eq.(17) one obtains
%%%%%%%%%%%%%%%%%%%%%%%%%%%%%%%%%%%%%%%%%%%%%%%%%%%%%%%%%%%%%%%%%%%%%%%%%%%%%
\begin{equation}
\Omega^{-1}(\tilde t)= \cosh(\zeta\tilde\phi(\tilde t)),
\end{equation}
%%%%%%%%%%%%%%%%%%%%%%%%%%%%%%%%%%%%%%%%%%%%%%%%%%%%%%%%%%%%%%%%%%%%%%%%%%%%%
and eq.(21) can be rewritten as
%%%%%%%%%%%%%%%%%%%%%%%%%%%%%%%%%%%%%%%%%%%%%%%%%%%%%%%%%%%%%%%%%%%%%%%%%%%%%
\begin{equation}
V(\phi)=(1-\zeta^{2}\phi^{2})^{2}\tilde V\left[\frac{\zeta^{-1}}{2}\ln
\left(\frac{1+\zeta\phi}{1-\zeta\phi}\right)\right].
\end{equation}
%%%%%%%%%%%%%%%%%%%%%%%%%%%%%%%%%%%%%%%%%%%%%%%%%%%%%%%%%%%%%%%%%%%%%%%%%%%%%
{}From the point of view of the search of exact solutions to the field
equations (3)--(5) in the conformal coupling case, this
changing frame process can be interpreted in the following way: if the set
$(\tilde g_{ab},\tilde \phi,\tilde V(\tilde \phi))$ forms a solution to the
field equations in the Einstein frame, that is, a solution to eqs.(3)--(5)
but with $\xi=0$, then the set $(g_{ab},\phi,V(\phi))$, where
$g_{ab}$, $\phi$ and $V(\phi)$  are given, respectively, by eqs.(16), (24)
and (26), with $\Omega$ given by eq.(25), constitutes a solution to the
same field equations but in the physical frame. The massless case of this
solution-generating technique was given by Bekenstein [14].

Note that this solution-generating technique is valid for any metric $g_{ab}$.
In the particular case that concerns us here, $g_{ab}$ is a FRW metric, whose
line element was given in eq.(6). The line element of the conformal
spacetime is, according to eq.(16), given by
%%%%%%%%%%%%%%%%%%%%%%%%%%%%%%%%%%%%%%%%%%%%%%%%%%%%%%%%%%%%%%%%%%%%%%%%%%%%%
\begin{equation}
d\tilde s^{2}=\Omega^{2}ds^{2}.
\end{equation}
%%%%%%%%%%%%%%%%%%%%%%%%%%%%%%%%%%%%%%%%%%%%%%%%%%%%%%%%%%%%%%%%%%%%%%%%%%%%%
This can be brought into the same form as eq.(6) by the following change of
variables
%%%%%%%%%%%%%%%%%%%%%%%%%%%%%%%%%%%%%%%%%%%%%%%%%%%%%%%%%%%%%%%%%%%%%%%%%%%%%
\begin{equation}
d\tilde t=\Omega dt,
\end{equation}
%%%%%%%%%%%%%%%%%%%%%%%%%%%%%%%%%%%%%%%%%%%%%%%%%%%%%%%%%%%%%%%%%%%%%%%%%%%%%
%%%%%%%%%%%%%%%%%%%%%%%%%%%%%%%%%%%%%%%%%%%%%%%%%%%%%%%%%%%%%%%%%%%%%%%%%%%%%
\begin{equation}
\tilde S=\Omega S.
\end{equation}
%%%%%%%%%%%%%%%%%%%%%%%%%%%%%%%%%%%%%%%%%%%%%%%%%%%%%%%%%%%%%%%%%%%%%%%%%%%%%
Then, we have in the Einstein frame
%%%%%%%%%%%%%%%%%%%%%%%%%%%%%%%%%%%%%%%%%%%%%%%%%%%%%%%%%%%%%%%%%%%%%%%%%%%%%
\begin{equation}
d\tilde s^{2}=-d\tilde t^{2}+\tilde S^{2}(\tilde t)\left(\frac{dr^{2}}
{1-kr^{2}}+r^{2}(d\theta^{2}+\sin^{2}\theta d\varphi^{2})\right),
\end{equation}
%%%%%%%%%%%%%%%%%%%%%%%%%%%%%%%%%%%%%%%%%%%%%%%%%%%%%%%%%%%%%%%%%%%%%%%%%%%%%
i.e., the metric in the Einstein frame $\tilde g_{ab}$ is also of the FRW
type, but with a redefined scale factor.

Before proceeding further, it will be useful to pause in order to introduce
some observational quantities of interest. The cosmological density parameter
$\Omega_{d}$ (we adopt this notation slightly at variance with the standard
usage to avoid any confusion with the conformal factor $\Omega$ ) is defined
by
%%%%%%%%%%%%%%%%%%%%%%%%%%%%%%%%%%%%%%%%%%%%%%%%%%%%%%%%%%%%%%%%%%%%%%%%%%%%%
\begin{equation}
\Omega_{d}=\frac{\rho_{\phi}}{3H^{2}},
\end{equation}
%%%%%%%%%%%%%%%%%%%%%%%%%%%%%%%%%%%%%%%%%%%%%%%%%%%%%%%%%%%%%%%%%%%%%%%%%%%%%
and when combined with the Friedmann equation (7), can be rewritten as
%%%%%%%%%%%%%%%%%%%%%%%%%%%%%%%%%%%%%%%%%%%%%%%%%%%%%%%%%%%%%%%%%%%%%%%%%%%%%
\begin{equation}
\Omega_{d}=1+\frac{k}{\dot{S}^{2}}.
\end{equation}
%%%%%%%%%%%%%%%%%%%%%%%%%%%%%%%%%%%%%%%%%%%%%%%%%%%%%%%%%%%%%%%%%%%%%%%%%%%%%
The deceleration parameter q is defined by
%%%%%%%%%%%%%%%%%%%%%%%%%%%%%%%%%%%%%%%%%%%%%%%%%%%%%%%%%%%%%%%%%%%%%%%%%%%%%
\begin{equation}
q=-\left(\frac{\ddot{S}}{H^{2}S}\right).
\end{equation}
%%%%%%%%%%%%%%%%%%%%%%%%%%%%%%%%%%%%%%%%%%%%%%%%%%%%%%%%%%%%%%%%%%%%%%%%%%%%%
Following [19], we take the sign of the deceleration parameter as indicating
whether a given cosmological model inflates or not. The positive sign
corresponds to ``standard" decelerating cosmological solutions whereas the
negative sign indicates inflation. Although there are several definitions
of what is actually meant by inflation, this is the most commonly used in
the study of exact scalar field cosmologies.

Defining a ``mixed" cosmic scale factor by $S(\tilde t)=S(t(\tilde t))$ we
have
%%%%%%%%%%%%%%%%%%%%%%%%%%%%%%%%%%%%%%%%%%%%%%%%%%%%%%%%%%%%%%%%%%%%%%%%%%%%%
\begin{equation}
\dot{S}=\Omega S^{\prime}
\end{equation}
%%%%%%%%%%%%%%%%%%%%%%%%%%%%%%%%%%%%%%%%%%%%%%%%%%%%%%%%%%%%%%%%%%%%%%%%%%%%%
%%%%%%%%%%%%%%%%%%%%%%%%%%%%%%%%%%%%%%%%%%%%%%%%%%%%%%%%%%%%%%%%%%%%%%%%%%%%%
\begin{equation}
\ddot{S}=\Omega(\Omega^{\prime} S^{\prime}+\Omega S^{\prime\prime}),
\end{equation}
%%%%%%%%%%%%%%%%%%%%%%%%%%%%%%%%%%%%%%%%%%%%%%%%%%%%%%%%%%%%%%%%%%%%%%%%%%%%%
where the prime denotes differentiation with respect to the cosmic time
$\tilde t$ of the Einstein frame. A very convenient form for $\Omega_{d}$
and $q$ can then be obtained inserting these results into eqs.(32) and (33)
%%%%%%%%%%%%%%%%%%%%%%%%%%%%%%%%%%%%%%%%%%%%%%%%%%%%%%%%%%%%%%%%%%%%%%%%%%%%%
\begin{equation}
\Omega_{d}=1+\frac{k}{\Omega^{2}S^{\prime 2}},
\end{equation}
%%%%%%%%%%%%%%%%%%%%%%%%%%%%%%%%%%%%%%%%%%%%%%%%%%%%%%%%%%%%%%%%%%%%%%%%%%%%%
%%%%%%%%%%%%%%%%%%%%%%%%%%%%%%%%%%%%%%%%%%%%%%%%%%%%%%%%%%%%%%%%%%%%%%%%%%%%%
\begin{equation}
q=-\frac{\Omega^{\prime} S}{\Omega S^{\prime}}-
\frac{S S^{\prime\prime}}{S^{\prime 2}}.
\end{equation}
%%%%%%%%%%%%%%%%%%%%%%%%%%%%%%%%%%%%%%%%%%%%%%%%%%%%%%%%%%%%%%%%%%%%%%%%%%%%%

For future reference, we end this section giving the equations of the
extended solution-generating technique for a spatially homogeneous
and isotropic metric (eqs.(24)--(26),(28)--(29)), rewritten in terms of
the $\psi$ scalar field variable
%%%%%%%%%%%%%%%%%%%%%%%%%%%%%%%%%%%%%%%%%%%%%%%%%%%%%%%%%%%%%%%%%%%%%%%%%%%%%
\begin{equation}
S(\tilde t)=\cosh(\tilde \psi(\tilde t))\tilde S(\tilde t),
\end{equation}
%%%%%%%%%%%%%%%%%%%%%%%%%%%%%%%%%%%%%%%%%%%%%%%%%%%%%%%%%%%%%%%%%%%%%%%%%%%%%
%%%%%%%%%%%%%%%%%%%%%%%%%%%%%%%%%%%%%%%%%%%%%%%%%%%%%%%%%%%%%%%%%%%%%%%%%%%%%
\begin{equation}
\frac{dt(\tilde t)}{d\tilde t}=\cosh(\tilde \psi(\tilde t)),
\end{equation}
%%%%%%%%%%%%%%%%%%%%%%%%%%%%%%%%%%%%%%%%%%%%%%%%%%%%%%%%%%%%%%%%%%%%%%%%%%%%%
%%%%%%%%%%%%%%%%%%%%%%%%%%%%%%%%%%%%%%%%%%%%%%%%%%%%%%%%%%%%%%%%%%%%%%%%%%%%%
\begin{equation}
\psi(\tilde t)=\tanh(\tilde \psi(\tilde t)),
\end{equation}
%%%%%%%%%%%%%%%%%%%%%%%%%%%%%%%%%%%%%%%%%%%%%%%%%%%%%%%%%%%%%%%%%%%%%%%%%%%%%
%%%%%%%%%%%%%%%%%%%%%%%%%%%%%%%%%%%%%%%%%%%%%%%%%%%%%%%%%%%%%%%%%%%%%%%%%%%%%
\begin{equation}
V(\psi)=(1-\psi^{2})^{2}\tilde V\left[\frac{\zeta^{-1}}{2}\ln\left
(\frac{1+\psi}{1-\psi}\right)\right].
\end{equation}
%%%%%%%%%%%%%%%%%%%%%%%%%%%%%%%%%%%%%%%%%%%%%%%%%%%%%%%%%%%%%%%%%%%%%%%%%%%%%
This equations can be looked upon as a map between the space of Einstein
frame solutions and the space of physical frame solutions.

\vspace{1.0cm}

{\Large\bf 4 \hspace{0.4cm} The Solutions}

\vspace{0.8cm}

We now apply the solution-generating technique developed in the preceding
section to a number of interesting situations. We consider, in the Einstein
frame, the solutions which were recently obtained by Ellis and Madsen [16].
Here only three of the five exact solutions found by Ellis and Madsen will
be mapped to the physical frame.

The titles adopted in the following subsections refer to the behaviour of
the solution in the Einstein frame.

\vspace{1.0cm}

{\large\bf 4.1 \hspace{0.4cm} Power-Law Expansion}

\vspace{0.8cm}

First, let us consider the solution in the Einstein frame characterised by
a power-law behaviour
%%%%%%%%%%%%%%%%%%%%%%%%%%%%%%%%%%%%%%%%%%%%%%%%%%%%%%%%%%%%%%%%%%%%%%%%%%%%%
\begin{equation}
\tilde{S}(\tilde{t})=A\tilde{t}^{n},
\end{equation}
%%%%%%%%%%%%%%%%%%%%%%%%%%%%%%%%%%%%%%%%%%%%%%%%%%%%%%%%%%%%%%%%%%%%%%%%%%%%%
%%%%%%%%%%%%%%%%%%%%%%%%%%%%%%%%%%%%%%%%%%%%%%%%%%%%%%%%%%%%%%%%%%%%%%%%%%%%%
\begin{equation}
\tilde{\psi}(\tilde{t})=\ln\tilde{t}^{p},
\end{equation}
%%%%%%%%%%%%%%%%%%%%%%%%%%%%%%%%%%%%%%%%%%%%%%%%%%%%%%%%%%%%%%%%%%%%%%%%%%%%%
%%%%%%%%%%%%%%%%%%%%%%%%%%%%%%%%%%%%%%%%%%%%%%%%%%%%%%%%%%%%%%%%%%%%%%%%%%%%%
\begin{equation}
\tilde{V}(\tilde{\psi})=V_{0}\exp\left(-\frac{2}{p}\tilde{\psi}\right),
\end{equation}
%%%%%%%%%%%%%%%%%%%%%%%%%%%%%%%%%%%%%%%%%%%%%%%%%%%%%%%%%%%%%%%%%%%%%%%%%%%%%
where $A, n$ are strictly positive constants with $n\neq 1$,
$p(n)=(n/3)^{1/2}, V_{0}(n)=n(3n-1)$, which is
only possible for $k=0$. This class of solutions is inflationary for
$n>1$, because if the power-law expansion continues for long enough, it will
solve all the well known problems of the standard hot big-bang model
(horizon, flatness, spectrum of primordial density perturbations,\ldots)[20].
However, it behaves like a ``standard" decelerating big-bang when
$0<n<1$.

Mapping the class of solutions (42)--(44) in the Einstein frame into the
physical frame with the help of eqs.(38)--(41), we obtain the following
parametric class of solutions in terms of the ``unphysical" time variable
$\tilde{t}$
%%%%%%%%%%%%%%%%%%%%%%%%%%%%%%%%%%%%%%%%%%%%%%%%%%%%%%%%%%%%%%%%%%%%%%%%%%%%%
\begin{equation}
S(\tilde{t})=\frac{A}{2}(\tilde{t}^{n+p}+\tilde{t}^{n-p}),
\end{equation}
%%%%%%%%%%%%%%%%%%%%%%%%%%%%%%%%%%%%%%%%%%%%%%%%%%%%%%%%%%%%%%%%%%%%%%%%%%%%%
%%%%%%%%%%%%%%%%%%%%%%%%%%%%%%%%%%%%%%%%%%%%%%%%%%%%%%%%%%%%%%%%%%%%%%%%%%%%%
\begin{equation}
t(\tilde{t})=\left\{ \begin{array}{ll}
\frac{1}{2(1+p)}\tilde{t}^{1+p}+\frac{1}{2(1-p)}\tilde{t}^{1-p} &
\mbox{if $p\neq 1$} \\ [2ex]
\frac{1}{4}\tilde{t}^{2}+\ln\tilde{t} & \mbox{if $p=1$} \end{array} \right.,
\end{equation}
%%%%%%%%%%%%%%%%%%%%%%%%%%%%%%%%%%%%%%%%%%%%%%%%%%%%%%%%%%%%%%%%%%%%%%%%%%%%%
%%%%%%%%%%%%%%%%%%%%%%%%%%%%%%%%%%%%%%%%%%%%%%%%%%%%%%%%%%%%%%%%%%%%%%%%%%%%%
\begin{equation}
\psi(\tilde{t})=\frac{\tilde{t}^{2p}-1}{\tilde{t}^{2p}+1},
\end{equation}
%%%%%%%%%%%%%%%%%%%%%%%%%%%%%%%%%%%%%%%%%%%%%%%%%%%%%%%%%%%%%%%%%%%%%%%%%%%%%
%%%%%%%%%%%%%%%%%%%%%%%%%%%%%%%%%%%%%%%%%%%%%%%%%%%%%%%%%%%%%%%%%%%%%%%%%%%%%
\begin{equation}
V(\psi)=V_{0}(1-\psi)^{\alpha}(1+\psi)^{\beta},
\end{equation}
%%%%%%%%%%%%%%%%%%%%%%%%%%%%%%%%%%%%%%%%%%%%%%%%%%%%%%%%%%%%%%%%%%%%%%%%%%%%%
where $\alpha(p)=2+1/p$ and $\beta(p)=2-1/p$. This class of
solutions is only consistent with a flat FRW geometry. As we shall find
in what follows, although it depends on the two parameters $A$ and $n$, its
behaviour is essentially determined by the $n$ parameter.

The cosmological density parameter for this class of solutions is
$\Omega_{d}=1$, while the deceleration parameter is given in terms of the
``unphysical" time variable $\tilde{t}$ by
%%%%%%%%%%%%%%%%%%%%%%%%%%%%%%%%%%%%%%%%%%%%%%%%%%%%%%%%%%%%%%%%%%%%%%%%%%%%%
\begin{equation}
q(\tilde{t})=\frac{q_{1}\tilde{t}^{4p}+q_{2}\tilde{t}^{2p}+q_{3}}
{(n+p)^{2}\tilde{t}^{4p}-q_{2}\tilde{t}^{2p}+(n-p)^{2}},
\end{equation}
%%%%%%%%%%%%%%%%%%%%%%%%%%%%%%%%%%%%%%%%%%%%%%%%%%%%%%%%%%%%%%%%%%%%%%%%%%%%%
where
%%%%%%%%%%%%%%%%%%%%%%%%%%%%%%%%%%%%%%%%%%%%%%%%%%%%%%%%%%%%%%%%%%%%%%%%%%%%%
\begin{equation}
q_{1}(n)=(n+p)(1-n),
\end{equation}
%%%%%%%%%%%%%%%%%%%%%%%%%%%%%%%%%%%%%%%%%%%%%%%%%%%%%%%%%%%%%%%%%%%%%%%%%%%%%
%%%%%%%%%%%%%%%%%%%%%%%%%%%%%%%%%%%%%%%%%%%%%%%%%%%%%%%%%%%%%%%%%%%%%%%%%%%%%
\begin{equation}
q_{2}(n)=2(p^{2}-n^{2}),
\end{equation}
%%%%%%%%%%%%%%%%%%%%%%%%%%%%%%%%%%%%%%%%%%%%%%%%%%%%%%%%%%%%%%%%%%%%%%%%%%%%%
%%%%%%%%%%%%%%%%%%%%%%%%%%%%%%%%%%%%%%%%%%%%%%%%%%%%%%%%%%%%%%%%%%%%%%%%%%%%%
\begin{equation}
q_{3}(n)=(n-p)(1-n).
\end{equation}
%%%%%%%%%%%%%%%%%%%%%%%%%%%%%%%%%%%%%%%%%%%%%%%%%%%%%%%%%%%%%%%%%%%%%%%%%%%%%

The behaviour of the class of solutions (45)--(48) is illustrated in
Figure 1 where $S(t)$ and $V(\psi)$ are plotted
for a representative set of values of the $A$,$n$ parameters.

Two points immediately stand out from eqs.(45) and (46). First, from
eq.(46) we can see that for $0<n<3$ we have
%%%%%%%%%%%%%%%%%%%%%%%%%%%%%%%%%%%%%%%%%%%%%%%%%%%%%%%%%%%%%%%%%%%%%%%%%%%%%
\begin{equation}
0<\tilde{t}<+\infty\quad{\Rightarrow}\quad 0<t<+\infty,
\end{equation}
%%%%%%%%%%%%%%%%%%%%%%%%%%%%%%%%%%%%%%%%%%%%%%%%%%%%%%%%%%%%%%%%%%%%%%%%%%%%%
while for $n\geq 3$ we have
%%%%%%%%%%%%%%%%%%%%%%%%%%%%%%%%%%%%%%%%%%%%%%%%%%%%%%%%%%%%%%%%%%%%%%%%%%%%%
\begin{equation}
0<\tilde{t}<+\infty\quad{\Rightarrow}\quad -\infty<t<+\infty.
\end{equation}
%%%%%%%%%%%%%%%%%%%%%%%%%%%%%%%%%%%%%%%%%%%%%%%%%%%%%%%%%%%%%%%%%%%%%%%%%%%%%
Secondly, from eq.(45) we can see that all the solutions of this class
corresponding to the range $\frac{1}{3}\leq n<3$ are big-bang solutions,
that is, start expanding from a Friedmann singularity $(S=0)$, while those
corresponding to the ranges $0<n<\frac{1}{3}$ and $n>3$ are singularity-free.
Indeed, for all solutions with $\frac{1}{3}\leq n<3$ the edge $S=0$ is always
reached in a finite proper time into the past
$(S\rightarrow 0$ as $t\rightarrow 0)$ (see Fig. 1(c)). On the other hand,
there is no Friedmann singularity for $0<n<\frac{1}{3}$
because these solutions always ``bounce" at a finite proper time into the
past given by (see Fig. 1(a))
%%%%%%%%%%%%%%%%%%%%%%%%%%%%%%%%%%%%%%%%%%%%%%%%%%%%%%%%%%%%%%%%%%%%%%%%%%%%%
\begin{equation}
t_{bounce}=\frac{1}{2(1+p)}\left(\frac{p-n}{n+p}\right)^{\frac{1+p}{2p}}+
\frac{1}{2(1-p)}\left(\frac{p-n}{n+p}\right)^{\frac{1-p}{2p}},
\end{equation}
%%%%%%%%%%%%%%%%%%%%%%%%%%%%%%%%%%%%%%%%%%%%%%%%%%%%%%%%%%%%%%%%%%%%%%%%%%%%%
neither for $n>3$ because, in this case, an infinite proper time into
the past is required to reach the edge $S=0$ (see Fig. 1(e)).

The asymptotic behaviour of the class of solutions (45)--(48) as
$\tilde{t}\rightarrow 0$ is given by
%%%%%%%%%%%%%%%%%%%%%%%%%%%%%%%%%%%%%%%%%%%%%%%%%%%%%%%%%%%%%%%%%%%%%%%%%%%%%
\begin{equation}
S(t)\sim \left\{ \begin{array}{ll}
t^{m} & \mbox{if $0<n<3$} \\[1ex]
(-t)^{m} & \mbox{ if $n>3$} \end{array} \right.,
\end{equation}
%%%%%%%%%%%%%%%%%%%%%%%%%%%%%%%%%%%%%%%%%%%%%%%%%%%%%%%%%%%%%%%%%%%%%%%%%%%%%
%%%%%%%%%%%%%%%%%%%%%%%%%%%%%%%%%%%%%%%%%%%%%%%%%%%%%%%%%%%%%%%%%%%%%%%%%%%%%
\begin{equation}
\psi(t)\sim \left\{ \begin{array}{ll}
-1+\psi_{0}t^{s} & \mbox{if $0<n<3$} \\[1ex]
-1+\psi_{0}(-t)^{s} & \mbox{if $n>3$} \end{array} \right. ,
\end{equation}
%%%%%%%%%%%%%%%%%%%%%%%%%%%%%%%%%%%%%%%%%%%%%%%%%%%%%%%%%%%%%%%%%%%%%%%%%%%%%
%%%%%%%%%%%%%%%%%%%%%%%%%%%%%%%%%%%%%%%%%%%%%%%%%%%%%%%%%%%%%%%%%%%%%%%%%%%%%
\begin{equation}
V(\psi)\sim (1+\psi)^{\beta},
\end{equation}
%%%%%%%%%%%%%%%%%%%%%%%%%%%%%%%%%%%%%%%%%%%%%%%%%%%%%%%%%%%%%%%%%%%%%%%%%%%%%
where
%%%%%%%%%%%%%%%%%%%%%%%%%%%%%%%%%%%%%%%%%%%%%%%%%%%%%%%%%%%%%%%%%%%%%%%%%%%%%
\begin{equation}
m(n)=\frac{n-p}{1-p}\quad, \quad s(n)=\frac{2p}{1-p},
\end{equation}
%%%%%%%%%%%%%%%%%%%%%%%%%%%%%%%%%%%%%%%%%%%%%%%%%%%%%%%%%%%%%%%%%%%%%%%%%%%%%
and
%%%%%%%%%%%%%%%%%%%%%%%%%%%%%%%%%%%%%%%%%%%%%%%%%%%%%%%%%%%%%%%%%%%%%%%%%%%%%
\begin{equation}
\psi_{0}=(2|1-p|)^{s},
\end{equation}
%%%%%%%%%%%%%%%%%%%%%%%%%%%%%%%%%%%%%%%%%%%%%%%%%%%%%%%%%%%%%%%%%%%%%%%%%%%%%
the asymptotic behaviour of the $n=3$ solution results from the above
equations in the limit $n\rightarrow 3$. Hence, we see that for an initial
inflationary phase to occur we must have
$m>1\Leftrightarrow 0<n<\frac{1}{3}$ or $n>1$. More precisely, the solutions
corresponding to the range $1<n<3$ exhibit an initial power-law
inflationary phase of expansion $(\dot{S}>0, \ddot{S}>0$ and $\dot{H}<0)$
which, in the limit $n\rightarrow 3$, yield to an initial phase of
standard De Sitter exponential inflation; those corresponding to the
range $n>3$ display an initial phase of pole inflation (superinflationary
expansion, $\dot{S}, \ddot{S}, \dot{H}$ all positive); finally, those
corresponding to the range $0<n<\frac{1}{3}$ exhibit an initial phase of
contraction characterised by $\dot{S}<0, \ddot{S}>0$ and $\dot{H}>0$,
which leads to a bounce. In this situation the criterion based on $q$
to define inflation, does not hold, since $\dot{S}<0$.

It is worth noticing that while in the minimal coupling case superinflation
only occurs for the $k=+1$ models [21], in the conformal coupling framework
that can occur for other values of the curvature index,
namely for $k=0$.

The late time behaviour $(\tilde{t}\rightarrow +\infty)$ of the class of
solutions (45)--(48) is given by
%%%%%%%%%%%%%%%%%%%%%%%%%%%%%%%%%%%%%%%%%%%%%%%%%%%%%%%%%%%%%%%%%%%%%%%%%%%%%
\begin{equation}
S(t)\sim t^{l},
\end{equation}
%%%%%%%%%%%%%%%%%%%%%%%%%%%%%%%%%%%%%%%%%%%%%%%%%%%%%%%%%%%%%%%%%%%%%%%%%%%%%
%%%%%%%%%%%%%%%%%%%%%%%%%%%%%%%%%%%%%%%%%%%%%%%%%%%%%%%%%%%%%%%%%%%%%%%%%%%%%
\begin{equation}
\psi(t)\sim 1-\psi_{\infty}t^{-r},
\end{equation}
%%%%%%%%%%%%%%%%%%%%%%%%%%%%%%%%%%%%%%%%%%%%%%%%%%%%%%%%%%%%%%%%%%%%%%%%%%%%%
%%%%%%%%%%%%%%%%%%%%%%%%%%%%%%%%%%%%%%%%%%%%%%%%%%%%%%%%%%%%%%%%%%%%%%%%%%%%%
\begin{equation}
V(\psi)\sim (1-\psi)^{\alpha},
\end{equation}
%%%%%%%%%%%%%%%%%%%%%%%%%%%%%%%%%%%%%%%%%%%%%%%%%%%%%%%%%%%%%%%%%%%%%%%%%%%%%
where
%%%%%%%%%%%%%%%%%%%%%%%%%%%%%%%%%%%%%%%%%%%%%%%%%%%%%%%%%%%%%%%%%%%%%%%%%%%%%
\begin{equation}
l(n)=\frac{n+p}{1+p} \quad , \quad r(n)=\frac{2p}{1+p},
\end{equation}
%%%%%%%%%%%%%%%%%%%%%%%%%%%%%%%%%%%%%%%%%%%%%%%%%%%%%%%%%%%%%%%%%%%%%%%%%%%%%
and
%%%%%%%%%%%%%%%%%%%%%%%%%%%%%%%%%%%%%%%%%%%%%%%%%%%%%%%%%%%%%%%%%%%%%%%%%%%%%
\begin{equation}
\psi_{\infty}=\left(2(1+p)\right)^{-r},
\end{equation}
%%%%%%%%%%%%%%%%%%%%%%%%%%%%%%%%%%%%%%%%%%%%%%%%%%%%%%%%%%%%%%%%%%%%%%%%%%%%%
so, we see that for a late time power-law inflationary phase to occur we
must have $l>1\Leftrightarrow n>1$, the limit $n\rightarrow +\infty$
corresponding to a late time standard De Sitter exponential inflationary
phase. While in the minimal coupling case
power-law inflation is normally driven by exponential potentials
$V(\psi)\propto \exp(-\lambda\psi)$, with $\lambda$ constant $>0$, we find
from eqs.(61),(63) that in the conformal coupling case power-law inflation
is driven by a polynomial potential.

All this conclusions on the asymptotic behaviour of the class of
solutions (45)--(48) are corroborated by a direct analysis of the asymptotic
behaviour of the deceleration parameter $q$ (cf. eq.(49)). Indeed, as
$\tilde{t}\rightarrow 0$ $q$ asymptotes to
%%%%%%%%%%%%%%%%%%%%%%%%%%%%%%%%%%%%%%%%%%%%%%%%%%%%%%%%%%%%%%%%%%%%%%%%%%%%%
\begin{equation}
q(t)\sim \frac{q_{3}}{(n-p)^{2}}=\frac{1-n}{n-p},
\end{equation}
%%%%%%%%%%%%%%%%%%%%%%%%%%%%%%%%%%%%%%%%%%%%%%%%%%%%%%%%%%%%%%%%%%%%%%%%%%%%%
yielding $q(t)<0$ only when $0<n<\frac{1}{3}$ or $n>1$, while its
late time behaviour $(\tilde{t}\rightarrow +\infty)$ is given by
%%%%%%%%%%%%%%%%%%%%%%%%%%%%%%%%%%%%%%%%%%%%%%%%%%%%%%%%%%%%%%%%%%%%%%%%%%%%%
\begin{equation}
q(t)\sim \frac{q_{1}}{(n+p)^{2}}=\frac{1-n}{n+p},
\end{equation}
%%%%%%%%%%%%%%%%%%%%%%%%%%%%%%%%%%%%%%%%%%%%%%%%%%%%%%%%%%%%%%%%%%%%%%%%%%%%%
yielding $q(t)<0$ only when $n>1$. Note that in these asymptotic
inflationary behaviours the slow-rolling approximation (see Ref.[23]) is
not valid.

An exceptional case occurs when $n=\frac{1}{3}$: on the one hand, the
potential is flat $V(\psi)=0$ and on the other, we can write
the corresponding solution explicitly in terms of the physical cosmic time t
%%%%%%%%%%%%%%%%%%%%%%%%%%%%%%%%%%%%%%%%%%%%%%%%%%%%%%%%%%%%%%%%%%%%%%%%%%%%%
\begin{equation}
\frac{S(t)}{A}=-\frac{1}{2}+\sqrt{\frac{4t+3}{12}},
\end{equation}
%%%%%%%%%%%%%%%%%%%%%%%%%%%%%%%%%%%%%%%%%%%%%%%%%%%%%%%%%%%%%%%%%%%%%%%%%%%%%
%%%%%%%%%%%%%%%%%%%%%%%%%%%%%%%%%%%%%%%%%%%%%%%%%%%%%%%%%%%%%%%%%%%%%%%%%%%%%
\begin{equation}
\psi(t)=1-2\sqrt{\frac{6}{16t+6}}.
\end{equation}
%%%%%%%%%%%%%%%%%%%%%%%%%%%%%%%%%%%%%%%%%%%%%%%%%%%%%%%%%%%%%%%%%%%%%%%%%%%%%
As $t\rightarrow +\infty$ the cosmic scale factor asymptotes to
%%%%%%%%%%%%%%%%%%%%%%%%%%%%%%%%%%%%%%%%%%%%%%%%%%%%%%%%%%%%%%%%%%%%%%%%%%%%%
\begin{equation}
S(t)\sim t^{1/2}.
\end{equation}
%%%%%%%%%%%%%%%%%%%%%%%%%%%%%%%%%%%%%%%%%%%%%%%%%%%%%%%%%%%%%%%%%%%%%%%%%%%%%
Thus, from a volumetric point of view, a flat FRW geometry filled with
a massless and non-interacting homogeneous conformally coupled scalar
field exhibits as late time behaviour that of the standard flat FRW
radiation-dominated solution. This should be compared with the analogous
situation in the Einstein frame, where a homogeneous free scalar field
propagating in a flat FRW background behaves as
%%%%%%%%%%%%%%%%%%%%%%%%%%%%%%%%%%%%%%%%%%%%%%%%%%%%%%%%%%%%%%%%%%%%%%%%%%%%%
\begin{equation}
S(t)\sim t^{1/3}.
\end{equation}
%%%%%%%%%%%%%%%%%%%%%%%%%%%%%%%%%%%%%%%%%%%%%%%%%%%%%%%%%%%%%%%%%%%%%%%%%%%%%

To summarize all these results on the global and asymptotic behaviour of
the class of solutions (45)--(48) we have collected them together in
Table 1.

\vspace{1.0cm}

{\large\bf 4.2 \hspace{0.4cm} Linear Expansion}

\vspace{0.8cm}

Next, consider the solution in the Einstein frame characterised by a linear
expansion behaviour
%%%%%%%%%%%%%%%%%%%%%%%%%%%%%%%%%%%%%%%%%%%%%%%%%%%%%%%%%%%%%%%%%%%%%%%%%%%%
\begin{equation}
\tilde{S}(\tilde{t})=A\tilde{t},
\end{equation}
%%%%%%%%%%%%%%%%%%%%%%%%%%%%%%%%%%%%%%%%%%%%%%%%%%%%%%%%%%%%%%%%%%%%%%%%%%%%
%%%%%%%%%%%%%%%%%%%%%%%%%%%%%%%%%%%%%%%%%%%%%%%%%%%%%%%%%%%%%%%%%%%%%%%%%%%%
\begin{equation}
\tilde{\psi}(\tilde{t})=\ln\tilde{t}^{p},
\end{equation}
%%%%%%%%%%%%%%%%%%%%%%%%%%%%%%%%%%%%%%%%%%%%%%%%%%%%%%%%%%%%%%%%%%%%%%%%%%%%
%%%%%%%%%%%%%%%%%%%%%%%%%%%%%%%%%%%%%%%%%%%%%%%%%%%%%%%%%%%%%%%%%%%%%%%%%%%%
\begin{equation}
\tilde{V}(\tilde{\psi})=V_{0}\exp\left(-\frac{2}{p}\tilde{\psi}\right),
\end{equation}
%%%%%%%%%%%%%%%%%%%%%%%%%%%%%%%%%%%%%%%%%%%%%%%%%%%%%%%%%%%%%%%%%%%%%%%%%%%%
where $A$ is a strictly positive constant,
%%%%%%%%%%%%%%%%%%%%%%%%%%%%%%%%%%%%%%%%%%%%%%%%%%%%%%%%%%%%%%%%%%%%%%%%%%%%
\begin{equation}
p(k,A)=\left(\frac{A^{2}+k}{3A^{2}}\right)^{1/2},
\end{equation}
%%%%%%%%%%%%%%%%%%%%%%%%%%%%%%%%%%%%%%%%%%%%%%%%%%%%%%%%%%%%%%%%%%%%%%%%%%%%
and $V_{0}(p)=\zeta^{-2}p^{2}$, which is always possible for $k=0,+1$
and can be made possible for $k=-1$ when subjected to the constraint $A>1$.
This class of coasting solutions may be viewed as a modern version of the
Milne solution (which is characterised by $p=\rho =0$ and $k=-1$ [17]), where
the scalar field matter source allows it to be non-empty and with a
generalization that any value of the curvature index $k$ is allowed [16].

Mapping the class of solutions (72)--(74) in the Einstein frame into the
physical frame with the help of eqs.(38)--(41), we obtain the following
parametric class of solutions in terms of the ``unphysical" time
variable $\tilde{t}$
%%%%%%%%%%%%%%%%%%%%%%%%%%%%%%%%%%%%%%%%%%%%%%%%%%%%%%%%%%%%%%%%%%%%%%%%%%%%
\begin{equation}
S(\tilde{t})=\frac{A}{2}(\tilde{t}^{1+p}+\tilde{t}^{1-p}),
\end{equation}
%%%%%%%%%%%%%%%%%%%%%%%%%%%%%%%%%%%%%%%%%%%%%%%%%%%%%%%%%%%%%%%%%%%%%%%%%%%%
%%%%%%%%%%%%%%%%%%%%%%%%%%%%%%%%%%%%%%%%%%%%%%%%%%%%%%%%%%%%%%%%%%%%%%%%%%%%
\begin{equation}
t(\tilde{t})=\left\{ \begin{array}{ll}
\frac{1}{2(1+p)}\tilde{t}^{1+p}+\frac{1}{2(1-p)}\tilde{t}^{1-p} &
\mbox{if $p\neq 1$} \\ [2ex]
\frac{1}{4}\tilde{t}^{2}+\frac{1}{2}\ln\tilde{t} &
\mbox{if $p=1$} \end{array} \right.,
\end{equation}
%%%%%%%%%%%%%%%%%%%%%%%%%%%%%%%%%%%%%%%%%%%%%%%%%%%%%%%%%%%%%%%%%%%%%%%%%%%%
%%%%%%%%%%%%%%%%%%%%%%%%%%%%%%%%%%%%%%%%%%%%%%%%%%%%%%%%%%%%%%%%%%%%%%%%%%%%
\begin{equation}
\psi(\tilde{t})=\frac{\tilde{t}^{2p}-1}{\tilde{t}^{2p}+1},
\end{equation}
%%%%%%%%%%%%%%%%%%%%%%%%%%%%%%%%%%%%%%%%%%%%%%%%%%%%%%%%%%%%%%%%%%%%%%%%%%%%
%%%%%%%%%%%%%%%%%%%%%%%%%%%%%%%%%%%%%%%%%%%%%%%%%%%%%%%%%%%%%%%%%%%%%%%%%%%%
\begin{equation}
V(\psi)=V_{0}(1-\psi)^{\alpha}(1+\psi)^{\beta},
\end{equation}
%%%%%%%%%%%%%%%%%%%%%%%%%%%%%%%%%%%%%%%%%%%%%%%%%%%%%%%%%%%%%%%%%%%%%%%%%%%%
where $\alpha(p)=2+1/p$ and $\beta(p)=2-1/p$. This class
of solutions also is consistent with all the FRW geometries, the
$A$-amplitude being in the $k=-1$ case subjected to the same constraint
as in the Einstein frame.

The cosmological density parameter for this class of solutions is given
in terms of the ``unphysical" time variable $\tilde{t}$ by
%%%%%%%%%%%%%%%%%%%%%%%%%%%%%%%%%%%%%%%%%%%%%%%%%%%%%%%%%%%%%%%%%%%%%%%%%%%%
\begin{equation}
\Omega_{d}(\tilde{t})=\frac{\omega_{1}\tilde{t}^{4p}+\omega_{2}\tilde{t}
^{2p}+\omega_{3}}{(\omega_{1}-k)\tilde{t}^{4p}+(\omega_{2}-2k)\tilde{t}
^{2p}+(\omega_{3}-k)},
\end{equation}
%%%%%%%%%%%%%%%%%%%%%%%%%%%%%%%%%%%%%%%%%%%%%%%%%%%%%%%%%%%%%%%%%%%%%%%%%%%%
where
%%%%%%%%%%%%%%%%%%%%%%%%%%%%%%%%%%%%%%%%%%%%%%%%%%%%%%%%%%%%%%%%%%%%%%%%%%%%
\begin{equation}
\omega_{1}(k,A)=A^{2}(1+p)^{2}+k,
\end{equation}
%%%%%%%%%%%%%%%%%%%%%%%%%%%%%%%%%%%%%%%%%%%%%%%%%%%%%%%%%%%%%%%%%%%%%%%%%%%%
%%%%%%%%%%%%%%%%%%%%%%%%%%%%%%%%%%%%%%%%%%%%%%%%%%%%%%%%%%%%%%%%%%%%%%%%%%%%
\begin{equation}
\omega_{2}(k,A)=2A^{2}(1-p^{2})+2k,
\end{equation}
%%%%%%%%%%%%%%%%%%%%%%%%%%%%%%%%%%%%%%%%%%%%%%%%%%%%%%%%%%%%%%%%%%%%%%%%%%%%
%%%%%%%%%%%%%%%%%%%%%%%%%%%%%%%%%%%%%%%%%%%%%%%%%%%%%%%%%%%%%%%%%%%%%%%%%%%%
\begin{equation}
\omega_{3}(k,A)=A^{2}(1-p)^{2}+k;
\end{equation}
%%%%%%%%%%%%%%%%%%%%%%%%%%%%%%%%%%%%%%%%%%%%%%%%%%%%%%%%%%%%%%%%%%%%%%%%%%%%
and the deceleration parameter is given by
%%%%%%%%%%%%%%%%%%%%%%%%%%%%%%%%%%%%%%%%%%%%%%%%%%%%%%%%%%%%%%%%%%%%%%%%%%%%
\begin{equation}
q(\tilde{t})=-\frac{4A^{2}p^{2}\tilde{t}^{2p}}
{(\omega_{1}-k)\tilde{t}^{4p}+(\omega_{2}-2k)\tilde{t}^{2p}+(\omega_{3}-k)}.
\end{equation}
%%%%%%%%%%%%%%%%%%%%%%%%%%%%%%%%%%%%%%%%%%%%%%%%%%%%%%%%%%%%%%%%%%%%%%%%%%%%

The behaviour of the cosmic density parameter of the class of solutions
(76)--(79), eq.(80), is illustrated in Figure 2 for closed
and open FRW geometries. For each of these cases
a representative set of values of the $p$ parameter is plotted.

Two points immediately
stand out from eqs.(76) and (77). First, from eq.(77) we can see that
for $0<p<1\Leftrightarrow A^{2}>k/2$ (this inequality is trivially
satisfied by all the flat and open solutions of the class, being also
satisfied by the closed solutions with $A^{2}>\frac{1}{2}$) we have
%%%%%%%%%%%%%%%%%%%%%%%%%%%%%%%%%%%%%%%%%%%%%%%%%%%%%%%%%%%%%%%%%%%%%%%%%%%%
\begin{equation}
0<\tilde{t}<+\infty \quad \Rightarrow \quad 0<t<+\infty
\end{equation}
%%%%%%%%%%%%%%%%%%%%%%%%%%%%%%%%%%%%%%%%%%%%%%%%%%%%%%%%%%%%%%%%%%%%%%%%%%%%
while for $p\geq 1\Leftrightarrow A^{2}\leq k/2$ (this inequality
is only satisfied by the closed solutions of the class with
$A^{2}\leq \frac{1}{2}$) we have
%%%%%%%%%%%%%%%%%%%%%%%%%%%%%%%%%%%%%%%%%%%%%%%%%%%%%%%%%%%%%%%%%%%%%%%%%%%%
\begin{equation}
0<\tilde{t}<+\infty \quad \Rightarrow \quad -\infty<t<+\infty.
\end{equation}
%%%%%%%%%%%%%%%%%%%%%%%%%%%%%%%%%%%%%%%%%%%%%%%%%%%%%%%%%%%%%%%%%%%%%%%%%%%%
Secondly, from eq.(76) we find that all the solutions of this class
corresponding to the range $0<p<1$ are big-bang solutions, while those
corresponding to the range $p\geq 1$ are singularity-free. There is no
Friedmann singularity for $p>1$ because these solutions always ``bounce"
at a finite proper time into the past (see Fig. 3(a)), given by
%%%%%%%%%%%%%%%%%%%%%%%%%%%%%%%%%%%%%%%%%%%%%%%%%%%%%%%%%%%%%%%%%%%%%%%%%%%%
\begin{equation}
t_{bounce}=\frac{1}{2(1+p)}\left(\frac{p-1}{p+1}\right)^{\frac{p+1}{2p}}
+\frac{1}{2(1-p)}\left(\frac{p-1}{p+1}\right)^{\frac{1-p}{2p}},
\end{equation}
%%%%%%%%%%%%%%%%%%%%%%%%%%%%%%%%%%%%%%%%%%%%%%%%%%%%%%%%%%%%%%%%%%%%%%%%%%%%
neither for $p=1$ because, in this case, the corresponding solution starts
expanding from a non-zero value of the cosmic scale factor $S_{\infty}$
(see Fig. 3(b)).

The asymptotic behaviour of the class of solutions (76)-(79) as
$\tilde{t}\rightarrow 0$ is given by
%%%%%%%%%%%%%%%%%%%%%%%%%%%%%%%%%%%%%%%%%%%%%%%%%%%%%%%%%%%%%%%%%%%%%%%%%%%%
\begin{equation}
S(t)\sim \left\{ \begin{array}{ll}
t & \mbox{if $p<1$} \\[1ex]
S_{\infty} & \mbox{if $p=1$} \\[1ex]
(-t) & \mbox{if $p>1$} \end{array} \right.,
\end{equation}
%%%%%%%%%%%%%%%%%%%%%%%%%%%%%%%%%%%%%%%%%%%%%%%%%%%%%%%%%%%%%%%%%%%%%%%%%%%%
%%%%%%%%%%%%%%%%%%%%%%%%%%%%%%%%%%%%%%%%%%%%%%%%%%%%%%%%%%%%%%%%%%%%%%%%%%%%
\begin{equation}
\psi(t)\sim \left\{ \begin{array}{ll}
-1+\psi_{0}t^{s} & \mbox{if $p<1$} \\[1ex]
-1+\exp(4t) & \mbox{if $p=1$} \\[1ex]
-1+\psi_{0}(-t)^{s} & \mbox{if $p>1$} \end{array} \right.,
\end{equation}
%%%%%%%%%%%%%%%%%%%%%%%%%%%%%%%%%%%%%%%%%%%%%%%%%%%%%%%%%%%%%%%%%%%%%%%%%%%%
%%%%%%%%%%%%%%%%%%%%%%%%%%%%%%%%%%%%%%%%%%%%%%%%%%%%%%%%%%%%%%%%%%%%%%%%%%%%
\begin{equation}
V(\psi)\sim (1+\psi)^{\beta},
\end{equation}
%%%%%%%%%%%%%%%%%%%%%%%%%%%%%%%%%%%%%%%%%%%%%%%%%%%%%%%%%%%%%%%%%%%%%%%%%%%%
where
%%%%%%%%%%%%%%%%%%%%%%%%%%%%%%%%%%%%%%%%%%%%%%%%%%%%%%%%%%%%%%%%%%%%%%%%%%%%
\begin{equation}
s(p)=\frac{2p}{1-p},
\end{equation}
%%%%%%%%%%%%%%%%%%%%%%%%%%%%%%%%%%%%%%%%%%%%%%%%%%%%%%%%%%%%%%%%%%%%%%%%%%%%
%%%%%%%%%%%%%%%%%%%%%%%%%%%%%%%%%%%%%%%%%%%%%%%%%%%%%%%%%%%%%%%%%%%%%%%%%%%%
\begin{equation}
\psi_{0}=(2|1-p|)^{s};
\end{equation}
%%%%%%%%%%%%%%%%%%%%%%%%%%%%%%%%%%%%%%%%%%%%%%%%%%%%%%%%%%%%%%%%%%%%%%%%%%%%
while the late time behaviour $(\tilde{t}\rightarrow +\infty)$ is given
by
%%%%%%%%%%%%%%%%%%%%%%%%%%%%%%%%%%%%%%%%%%%%%%%%%%%%%%%%%%%%%%%%%%%%%%%%%%%%
\begin{equation}
S(t)\sim t,
\end{equation}
%%%%%%%%%%%%%%%%%%%%%%%%%%%%%%%%%%%%%%%%%%%%%%%%%%%%%%%%%%%%%%%%%%%%%%%%%%%%
%%%%%%%%%%%%%%%%%%%%%%%%%%%%%%%%%%%%%%%%%%%%%%%%%%%%%%%%%%%%%%%%%%%%%%%%%%%%
\begin{equation}
\psi(t)\sim 1-\psi_{\infty}t^{-r},
\end{equation}
%%%%%%%%%%%%%%%%%%%%%%%%%%%%%%%%%%%%%%%%%%%%%%%%%%%%%%%%%%%%%%%%%%%%%%%%%%%%
%%%%%%%%%%%%%%%%%%%%%%%%%%%%%%%%%%%%%%%%%%%%%%%%%%%%%%%%%%%%%%%%%%%%%%%%%%%%
\begin{equation}
V(\psi)\sim (1-\psi)^{\alpha},
\end{equation}
%%%%%%%%%%%%%%%%%%%%%%%%%%%%%%%%%%%%%%%%%%%%%%%%%%%%%%%%%%%%%%%%%%%%%%%%%%%%%
where
%%%%%%%%%%%%%%%%%%%%%%%%%%%%%%%%%%%%%%%%%%%%%%%%%%%%%%%%%%%%%%%%%%%%%%%%%%%%%
\begin{equation}
r(p)=\frac{2p}{1-p},
\end{equation}
%%%%%%%%%%%%%%%%%%%%%%%%%%%%%%%%%%%%%%%%%%%%%%%%%%%%%%%%%%%%%%%%%%%%%%%%%%%%%
%%%%%%%%%%%%%%%%%%%%%%%%%%%%%%%%%%%%%%%%%%%%%%%%%%%%%%%%%%%%%%%%%%%%%%%%%%%%%
\begin{equation}
\psi_{\infty}=\left(2(1+p)\right)^{-r}.
\end{equation}
%%%%%%%%%%%%%%%%%%%%%%%%%%%%%%%%%%%%%%%%%%%%%%%%%%%%%%%%%%%%%%%%%%%%%%%%%%%%%
Hence, we see that the class of solutions (76)--(79) does not exhibit an
asymptotic inflationary phase neither as $\tilde{t}\rightarrow 0$ nor as
$\tilde{t}\rightarrow +\infty$ approaching instead, in both cases and for
general $p$, the $S(t)\sim t$ behaviour of the coasting solutions. This
conclusion is corroborated by a direct analysis of the asymptotic
behaviour of the deceleration parameter since $q\rightarrow 0$ both
as $\tilde{t}\rightarrow 0$ and as $\tilde{t}\rightarrow +\infty$
(cf. eq.(84)).

Although the class of solutions (76)-(79) is not asymptotically inflationary
according to the negative deceleration parameter criterion, it has been
argued that the coasting solutions ought to be considered inflationary [21].
In fact, if the linear expansion continues for long enough it will solve
all the well known kinematical problems of the standard hot big-bang model.
Furthermore, as shown in Ref. [22], these models allow the generation
and evolution of density perturbations that may be responsible for the
large scale structure that we observe in the Universe. However, unlike what
is usually expected from inflationary models the present models
can lead to $\Omega_{d}\neq 1$ today [21]. This interesting feature of the
coasting models, which may provide a solution to the cosmic dark matter
problem [23], can be illustrated by the asymptotic behaviour of the
cosmological density parameter corresponding to the class of solutions
(76)--(79), since as $\tilde{t}\rightarrow 0$ we have (see Fig. 4)
%%%%%%%%%%%%%%%%%%%%%%%%%%%%%%%%%%%%%%%%%%%%%%%%%%%%%%%%%%%%%%%%%%%%%%%%%%%%%
\begin{equation}
\Omega_{d}(t)\sim\frac{\omega_{3}}{\omega_{3}-k}=1+\frac{k}
{A^{2}(1-p)^{2}},
\end{equation}
%%%%%%%%%%%%%%%%%%%%%%%%%%%%%%%%%%%%%%%%%%%%%%%%%%%%%%%%%%%%%%%%%%%%%%%%%%%%%
and as $\tilde{t}\rightarrow +\infty$ we have
%%%%%%%%%%%%%%%%%%%%%%%%%%%%%%%%%%%%%%%%%%%%%%%%%%%%%%%%%%%%%%%%%%%%%%%%%%%%%
\begin{equation}
\Omega_{d}(t)\sim\frac{\omega_{1}}{\omega_{1}-k}=1+\frac{k}
{A^{2}(1+p)^{2}}.
\end{equation}
%%%%%%%%%%%%%%%%%%%%%%%%%%%%%%%%%%%%%%%%%%%%%%%%%%%%%%%%%%%%%%%%%%%%%%%%%%%%%

\vspace{1.0cm}

{\large\bf 4.3 \hspace{0.4cm} De Sitter Expansion From a Singularity}

\vspace{0.8cm}

Finally, let us consider the solution in the Einstein frame characterised
by a ``$\sinh$" expansion from a Friedmann singularity $(\tilde{S}=0)$
%%%%%%%%%%%%%%%%%%%%%%%%%%%%%%%%%%%%%%%%%%%%%%%%%%%%%%%%%%%%%%%%%%%%%%%%%%%%%
\begin{equation}
\tilde{S}(\tilde{t})=A\sinh(\omega\tilde{t}),
\end{equation}
%%%%%%%%%%%%%%%%%%%%%%%%%%%%%%%%%%%%%%%%%%%%%%%%%%%%%%%%%%%%%%%%%%%%%%%%%%%%%
%%%%%%%%%%%%%%%%%%%%%%%%%%%%%%%%%%%%%%%%%%%%%%%%%%%%%%%%%%%%%%%%%%%%%%%%%%%%%
\begin{equation}
\tilde{\psi}(\tilde{t})=\ln\left(\tanh^{p}(\frac{\omega\tilde{t}}{2})\right),
\end{equation}
%%%%%%%%%%%%%%%%%%%%%%%%%%%%%%%%%%%%%%%%%%%%%%%%%%%%%%%%%%%%%%%%%%%%%%%%%%%%%
%%%%%%%%%%%%%%%%%%%%%%%%%%%%%%%%%%%%%%%%%%%%%%%%%%%%%%%%%%%%%%%%%%%%%%%%%%%%%
\begin{equation}
\tilde{V}(\tilde{\psi})=V_{0}\left(\frac{1}{2}+p^{2}\sinh^{2}(\frac{1}{p}
\tilde{\psi})\right),
\end{equation}
%%%%%%%%%%%%%%%%%%%%%%%%%%%%%%%%%%%%%%%%%%%%%%%%%%%%%%%%%%%%%%%%%%%%%%%%%%%%%
where $A,\omega$ are strictly positive constants,
%%%%%%%%%%%%%%%%%%%%%%%%%%%%%%%%%%%%%%%%%%%%%%%%%%%%%%%%%%%%%%%%%%%%%%%%%%%%%
\begin{equation}
p(\omega,A)=\left(\frac{1}{3}+\frac{k}{3A^{2}\omega^{2}}\right)^{1/2},
\end{equation}
%%%%%%%%%%%%%%%%%%%%%%%%%%%%%%%%%%%%%%%%%%%%%%%%%%%%%%%%%%%%%%%%%%%%%%%%%%%%%
and $V_{0}=\zeta^{-2}\omega^{2}$, which is always possible for $k=0,+1$
and can be made possible for $k=-1$ when subjected to the constraint
$A\omega>1$. For large $\tilde{t}$, this class
of solutions approaches a standard De Sitter inflationary phase, the
cosmic scale factor growing exponentially with time in this limit.

Mapping the class of solutions (100)--(102) in the Einstein frame into
the physical frame with the help of eqs.(38)--(41), we obtain the following
parametric class of solutions in terms of the ``unphysical" time
variable $\tilde{t}$
%%%%%%%%%%%%%%%%%%%%%%%%%%%%%%%%%%%%%%%%%%%%%%%%%%%%%%%%%%%%%%%%%%%%%%%%%%%%%
\begin{equation}
S(\tilde{t})=\frac{A}{2}\sinh(\omega\tilde{t})\left(\tanh^{p}
(\frac{\omega\tilde{t}}{2})+\coth^{p}(\frac{\omega\tilde{t}}{2})\right),
\end{equation}
%%%%%%%%%%%%%%%%%%%%%%%%%%%%%%%%%%%%%%%%%%%%%%%%%%%%%%%%%%%%%%%%%%%%%%%%%%%%%
%%%%%%%%%%%%%%%%%%%%%%%%%%%%%%%%%%%%%%%%%%%%%%%%%%%%%%%%%%%%%%%%%%%%%%%%%%%%%
\begin{equation}
\frac{dt(\tilde{t})}{d\tilde{t}}=\frac{1}{2}
\left(\tanh^{p}(\frac{\omega\tilde{t}}{2})+
\coth^{p}(\frac{\omega\tilde{t}}{2})\right),
\end{equation}
%%%%%%%%%%%%%%%%%%%%%%%%%%%%%%%%%%%%%%%%%%%%%%%%%%%%%%%%%%%%%%%%%%%%%%%%%%%%%
%%%%%%%%%%%%%%%%%%%%%%%%%%%%%%%%%%%%%%%%%%%%%%%%%%%%%%%%%%%%%%%%%%%%%%%%%%%%%
\begin{equation}
\psi(\tilde{t})=\frac{\tanh^{2p}(\frac{\omega\tilde{t}}{2})-1}
{\tanh^{2p}(\frac{\omega\tilde{t}}{2})+1},
\end{equation}
%%%%%%%%%%%%%%%%%%%%%%%%%%%%%%%%%%%%%%%%%%%%%%%%%%%%%%%%%%%%%%%%%%%%%%%%%%%%%
%%%%%%%%%%%%%%%%%%%%%%%%%%%%%%%%%%%%%%%%%%%%%%%%%%%%%%%%%%%%%%%%%%%%%%%%%%%%%
\begin{equation}
V(\psi)=V_{1}(1-\psi^{2})^{2}+V_{2}\left((1+\psi)^{\alpha}(1-\psi)^{\beta}
+(1+\psi)^{\beta}(1-\psi)^{\alpha}\right),
\end{equation}
%%%%%%%%%%%%%%%%%%%%%%%%%%%%%%%%%%%%%%%%%%%%%%%%%%%%%%%%%%%%%%%%%%%%%%%%%%%%%
where
%%%%%%%%%%%%%%%%%%%%%%%%%%%%%%%%%%%%%%%%%%%%%%%%%%%%%%%%%%%%%%%%%%%%%%%%%%%%%
\begin{equation}
\alpha(p)=2+\frac{1}{p} \quad , \quad \beta(p)=2-\frac{1}{p},
\end{equation}
%%%%%%%%%%%%%%%%%%%%%%%%%%%%%%%%%%%%%%%%%%%%%%%%%%%%%%%%%%%%%%%%%%%%%%%%%%%%%
and
%%%%%%%%%%%%%%%%%%%%%%%%%%%%%%%%%%%%%%%%%%%%%%%%%%%%%%%%%%%%%%%%%%%%%%%%%%%%%
\begin{equation}
V_{1}(\omega,A)=2\omega^{2}-\frac{1}{A^{2}} \quad , \quad V_{2}
(\omega,A)=\frac{1}{2}\left(\omega^{2}+\frac{1}{A^{2}}\right).
\end{equation}
%%%%%%%%%%%%%%%%%%%%%%%%%%%%%%%%%%%%%%%%%%%%%%%%%%%%%%%%%%%%%%%%%%%%%%%%%%%%%
Here $p$ is restricted to be a positive integer and $A,\omega$ are
fixed in such a way as to satisfy the following constraint equation
%%%%%%%%%%%%%%%%%%%%%%%%%%%%%%%%%%%%%%%%%%%%%%%%%%%%%%%%%%%%%%%%%%%%%%%%%%%%%
\begin{equation}
A^{2}\omega^{2}=\frac{1}{3p^{2}-1}.
\end{equation}
%%%%%%%%%%%%%%%%%%%%%%%%%%%%%%%%%%%%%%%%%%%%%%%%%%%%%%%%%%%%%%%%%%%%%%%%%%%%%

Although the starting class of solutions in the Einstein frame eqs.(100)--(102)
is consistent with all the FRW geometries, this class is only
consistent with a closed FRW geometry. This reduction in the number of
admissible FRW geometric backgrounds, is a consequence of the integration
of eq.(105) only being possible when $p$ is a positive integer, which
happens only when $k=+1$. However, the integrals of integer powers
of the ``$\tanh$" and ``$\coth$" functions are given inductively [24] so,
we can not write the solution of eq.(105) in closed form for general
$p$. To accomplish this we have to restrict ourselves to a particular
value of $p$.

In what follows we will focus, for simplicity, on the $p=1$ solution.
Taking $p=1$ in eqs.(104)--(107), we get
%%%%%%%%%%%%%%%%%%%%%%%%%%%%%%%%%%%%%%%%%%%%%%%%%%%%%%%%%%%%%%%%%%%%%%%%%%%%%
\begin{equation}
S(\tilde{t})=A\cosh(\omega\tilde{t}),
\end{equation}
%%%%%%%%%%%%%%%%%%%%%%%%%%%%%%%%%%%%%%%%%%%%%%%%%%%%%%%%%%%%%%%%%%%%%%%%%%%%%
%%%%%%%%%%%%%%%%%%%%%%%%%%%%%%%%%%%%%%%%%%%%%%%%%%%%%%%%%%%%%%%%%%%%%%%%%%%%%
\begin{equation}
t(\tilde{t})=\frac{1}{\omega}\ln(\sinh(\omega\tilde{t})),
\end{equation}
%%%%%%%%%%%%%%%%%%%%%%%%%%%%%%%%%%%%%%%%%%%%%%%%%%%%%%%%%%%%%%%%%%%%%%%%%%%%%
%%%%%%%%%%%%%%%%%%%%%%%%%%%%%%%%%%%%%%%%%%%%%%%%%%%%%%%%%%%%%%%%%%%%%%%%%%%%%
\begin{equation}
\psi^{-1}(\tilde{t})=-\cosh(\omega\tilde{t}),
\end{equation}
%%%%%%%%%%%%%%%%%%%%%%%%%%%%%%%%%%%%%%%%%%%%%%%%%%%%%%%%%%%%%%%%%%%%%%%%%%%%%
%%%%%%%%%%%%%%%%%%%%%%%%%%%%%%%%%%%%%%%%%%%%%%%%%%%%%%%%%%%%%%%%%%%%%%%%%%%%%
\begin{equation}
V(\psi)=V_{0}(1-\psi^{4}),
\end{equation}
%%%%%%%%%%%%%%%%%%%%%%%%%%%%%%%%%%%%%%%%%%%%%%%%%%%%%%%%%%%%%%%%%%%%%%%%%%%%%
where $A^{2}\omega^{2}=\frac{1}{2}$ and $V_{0}=3/2A^{2}$.

Now, it's interesting to note that eq.(112) can be solved for $t$,
which enables one to eliminate the $\tilde{t}$ parameter between
eqs.(111)--(113) and to write the solution explicitly in terms of the
physical cosmic time $t$. When this is done, we obtain
%%%%%%%%%%%%%%%%%%%%%%%%%%%%%%%%%%%%%%%%%%%%%%%%%%%%%%%%%%%%%%%%%%%%%%%%%%%%%
\begin{equation}
S(t)=A(1+\exp(2\omega t))^{1/2},
\end{equation}
%%%%%%%%%%%%%%%%%%%%%%%%%%%%%%%%%%%%%%%%%%%%%%%%%%%%%%%%%%%%%%%%%%%%%%%%%%%%%
%%%%%%%%%%%%%%%%%%%%%%%%%%%%%%%%%%%%%%%%%%%%%%%%%%%%%%%%%%%%%%%%%%%%%%%%%%%%%
\begin{equation}
\psi(t)=-(1+\exp(2\omega t))^{-1/2},
\end{equation}
%%%%%%%%%%%%%%%%%%%%%%%%%%%%%%%%%%%%%%%%%%%%%%%%%%%%%%%%%%%%%%%%%%%%%%%%%%%%%
with $V(\psi)$ given by eq.(114).

The cosmological density parameter for this solution is given in terms
of the physical cosmic time $t$ by
%%%%%%%%%%%%%%%%%%%%%%%%%%%%%%%%%%%%%%%%%%%%%%%%%%%%%%%%%%%%%%%%%%%%%%%%%%%%%
\begin{equation}
\Omega_{d}(t)=1+2(\exp(-2\omega t)+\exp(-4\omega t)),
\end{equation}
%%%%%%%%%%%%%%%%%%%%%%%%%%%%%%%%%%%%%%%%%%%%%%%%%%%%%%%%%%%%%%%%%%%%%%%%%%%%%
and the deceleration parameter is given by
%%%%%%%%%%%%%%%%%%%%%%%%%%%%%%%%%%%%%%%%%%%%%%%%%%%%%%%%%%%%%%%%%%%%%%%%%%%%%
\begin{equation}
q(t)=-\left(\frac{1}{\omega^{2}}+\exp(-2\omega t)\right).
\end{equation}
%%%%%%%%%%%%%%%%%%%%%%%%%%%%%%%%%%%%%%%%%%%%%%%%%%%%%%%%%%%%%%%%%%%%%%%%%%%%%

For very large $t$ $(t\rightarrow +\infty)$, the cosmic scale factor
asymptotes to
%%%%%%%%%%%%%%%%%%%%%%%%%%%%%%%%%%%%%%%%%%%%%%%%%%%%%%%%%%%%%%%%%%%%%%%%%%%%
\begin{equation}
S(t)\sim \exp(\omega t),
\end{equation}
%%%%%%%%%%%%%%%%%%%%%%%%%%%%%%%%%%%%%%%%%%%%%%%%%%%%%%%%%%%%%%%%%%%%%%%%%%%%%
while for $t\rightarrow -\infty$ we have
%%%%%%%%%%%%%%%%%%%%%%%%%%%%%%%%%%%%%%%%%%%%%%%%%%%%%%%%%%%%%%%%%%%%%%%%%%%%%
\begin{equation}
S(t)\sim A.
\end{equation}
%%%%%%%%%%%%%%%%%%%%%%%%%%%%%%%%%%%%%%%%%%%%%%%%%%%%%%%%%%%%%%%%%%%%%%%%%%%%%
This is a non-singular inflationary solution that approaches asymptotically
a standard De Sitter exponential inflationary phase. Such late time
inflationary phase results here, as in the minimal coupling case, from the
presence of the constant term $V_{0}$ in eq.(114),
which is just an effective cosmological constant. On the other hand,
the cosmological density parameter makes a smooth transition from infinity
$(\Omega_{d}\rightarrow +\infty$ as $t\rightarrow -\infty)$ to unity
$(\Omega_{d}\rightarrow 1$ as $t\rightarrow +\infty)$, so the solution
is asymptotically flat.

It is worth noticing that the slow-rolling condition (see Ref.[23])
although asymptotically satisfied in the Einstein frame, fails to be true
in the physical frame. The ratio $\Delta=|\dot{\psi}/\psi|/H$ is equal to
unity instead of $\Delta\ll 1$. So, to a slow-rolling
inflationary phase in the Einstein frame do not always corresponds
a slow-rolling inflationary phase in the original frame. In fact, the
condition for this to happen requires $\psi^{2}\ll 1$ as shown in Ref.[8],
which is violated in our case $(\psi^{2}<1)$.

\vspace{1.0cm}

{\Large\bf 5 \hspace{0.4cm} Summary and Conclusions}

\vspace{0.8cm}

In this work we have found a number of new, exact classes of conformal
scalar field cosmologies. The solution-generating technique used here
extends the technique originally introduced by Bekenstein for massless
conformal scalar fields [14], and is based on the conformal
equivalence between the non-minimal coupling model characterised by the
action (1) and canonical Einstein's gravity plus a canonical scalar field.
This technique has been  used in the literature to map a number of
familiar results from the Einstein frame, such as the reheating temperature
and density spectrum, into the physical frame [25]. It has also been used to
derive new exact cosmological solutions to some scalar-tensor theories [26].
Nevertheless, to our knowledge, the conformal transformation technique
has never been used to derive self-interacting conformal scalar
field cosmologies.

We have applied the extended solution-generating technique to the
classes of solutions in the Einstein frame that were recently obtained
by Ellis and Madsen [16]. For the classes of solutions obtained
in the physical frame we have discussed the existence of singularities,
the asymptotic behaviour and the possibility of occurrence of
inflationary phases. As a result, we have found that one class of
solutions (eqs.(45)--(48)) tends asymptotically to a power-law
inflationary behaviour $S(t)\sim t^{p}$ with $p>1$ yielding, in the limit
$p\rightarrow +\infty$, to a late time phase of standard De Sitter
exponential inflation; and that another class (eqs.(76)--(79)) exhibits
a late time approach to the $S(t)\sim t$ behaviour of the coasting models,
leading to any asymptotic value for the cosmological density parameter, in
particular to $\Omega_{d}\neq 1$, depending on the choice of the initial
conditions. Both of these classes comprise bouncing solutions which avoid
an initial singularity. It was also found that, a flat FRW geometry filled
with a massless and non-interacting homogeneous conformally coupled scalar
field approaches asymptotically the typical $S(t)\sim t^{1/2}$ behaviour
of the standard flat FRW radiation-dominated solution. Our results thus
reveal that to simple behaviours of the cosmic scale factor in the
Einstein frame corresponds a wealth of behaviours in the physical frame,
namely, we find several different phases, with in some cases, a smooth
transition between ordinary and inflationary expansion. Note that to a
slow-rolling inflationary phase in the Einstein frame do not always
corresponds a slow-rolling inflationary phase in the physical frame. This
indicates that some caution is required in transposing to the physical
frame results which are derived in the Einstein frame, in particular when
regarding inflation.

This work also serves the purpose of establishing the classes of potentials
in the physical frame which drive the various behaviours of the cosmic scale
factor considered in the Einstein frame. This allows a comparison to be made
with the corresponding potentials in the Einstein frame. For instance, while
in the minimal coupling case a power-law inflationary phase is
normally driven by exponential potentials $V(\psi)\propto\exp(-\lambda\psi)$,
with $\lambda$ constant $>0$, we have found that in the conformal coupling
case it is driven by a polynomial potential(cf. eq.(63)).

Although we have focused on some classes of solutions in the Einstein
frame, the technique used here can also be utilised to generate other
conformal scalar field cosmologies by starting from other  minimal
scalar field cosmologies. Furthermore, the extented solution-generating
technique can also be applied to other geometric backgrounds. In a
forthcoming paper we will apply it to a number of anisotropic scalar
field cosmologies.

\vspace{2cm}
{\Large\bf Acknowledgements}

The authors are grateful to Paulo Vargas Moniz for careful reading
an early draft of the manuscript and for very helpful comments. JPA was
supported by an JNICT graduate scholarship during this research
(Programa Ci\^{e}ncia BM/2054/91-RM). The exact solutions given in
this paper were verified using the symbolic computation package
MATHEMATICA and the graphics were produced with GLE 3.2.

\pagebreak

\pagebreak

\begin{center}
{\Large\bf Table Captions}
\end{center}

\hspace*{0.5cm}{\bf Table 1:}Summary of the global and asymptotic
volumetric behaviour of the solutions of the power-law expansion class
(eqs.(45)--(48)).

\vspace*{2.0cm}

\begin{center}
{\Large\bf Figure Captions}
\end{center}

\hspace*{0.5cm}{\bf Figure 1:}The figure shows the scale factor plotted
against the time $t$ and the potential plotted against $\psi$ for the
power-law expansion class (eqs.(45)--(48)):
(a)-(b) $n=\frac{1}{4}$ and $A=2$ $(\Rightarrow p=\sqrt{\frac{1}{12}})$,
(c)-(d) $n=\frac{1}{2}$ and $A=2$ $(\Rightarrow p=\sqrt{\frac{1}{6}})$,
(e)-(f) $n=12$ and $A=2$ $(\Rightarrow p=2)$.
The graphs (a),(c) and (e) correspond to the scale factor.
The graphs (b),(d) and (f) correspond to the potential.

\vspace*{1.0cm}

\hspace*{0.5cm}{\bf Figure 2:}The figure shows the time evolution of the
cosmological density parameter $\Omega_{d}$ for the linear expansion class
(eqs.(76)--(79)): (a) $k=+1$ and $p=\frac{2}{3}$ $(\Rightarrow A=\sqrt{3})$,
(b) $k=+1$ and $p=1$ $(\Rightarrow A=\frac{1}{\sqrt{2}})$,
(c) $k=+1$ and $p=2$ $(\Rightarrow A=\frac{1}{\sqrt{11}})$,
(d) $k=-1$ and $p=\sqrt{\frac{5}{27}}$ $(\Rightarrow A=\frac{3}{2})$.

\vspace*{1.0cm}

\hspace*{0.5cm}{\bf Figure 3:}The figure shows the scale factor plotted
against the time $t$ for the solutions of the linear expansion class
(eqs.(76)--(79)) corresponding to the parameter values
(a) $p=2$ $(\Rightarrow A=\sqrt{\frac{1}{11}})$ and
(b) $p=1$ $(\Rightarrow A=\sqrt{\frac{1}{6}})$.

\vspace*{1.0cm}

\hspace*{0.5cm}{\bf Figure 4:}The figure shows the asymptotic values
taken by the cosmological density parameter $\Omega_{d}$ as
$\tilde{t}\rightarrow 0$ $(\Omega_{0})$ and as
$\tilde{t}\rightarrow +\infty$ $(\Omega_{\infty})$ plotted against
the A parameter, for the solutions of the linear
expansion class (eqs.(76)--(79)). The graphs (a) and (b) correspond
to the closed $(k=+1)$ model. In (a) the graph diverges to
infinity as $A\rightarrow \sqrt{\frac{1}{2}}$, asymptotes to
$4$ as $A\rightarrow 0$ and to $1$ as $A\rightarrow +\infty$.
The graphs (c) and (d) correspond to the open $(k=-1)$ model.

\pagebreak
\vspace*{4cm}
\begin{center}
{\Large\bf Table 1}
\end{center}
\vspace{2cm}
{\footnotesize
\begin{tabular} {|c|c|c|c|} \hline\hline
   &{\em Global}&\multicolumn{2}{c|}{\em Asymptotic Behaviour} \\
\cline{3-4}
 &{\em Behaviour} & $\tilde{t}\rightarrow 0$ &
 $ \tilde{t}\rightarrow +\infty$ \\ \hline
$0<n<1/3$ & Bouncing Models & Contraction & Ordinary Expansion \\
 \hline
$1/3 \leq n<1$ & ``Standard" Big-Bang Models & Ordinary Expansion &
Ordinary Expansion \\  \hline
$n=1$ & ``Standard" Big-Bang Model & Linear Expansion & Linear Expansion \\
\hline
$1<n<3$ & Inflationary Big-Bang Models & Power-Law Inflation & Power-Law
Inflation \\ \hline
$n=3$ & Non-Singular Inflationary Model & De Sitter Expansion & Power-Law
Inflation \\ \hline
$n>3$ & Non-Singular Inflationary Models & Pole Inflation & Power-Law
Inflation \\ \hline
\hline
\end{tabular} }

\end{document}